\documentclass[aps, 10pt, prd, notitlepage, twocolumn,
superscriptaddress,nofootinbib, floatfix, tightenlines]{revtex4-1}

\usepackage{amsmath}
\usepackage{amssymb}
\usepackage{amsfonts}
\usepackage[utf8]{inputenc}
\usepackage[T1]{fontenc}
\usepackage{mathrsfs}
\usepackage{anyfontsize}
\usepackage{mathtools}
\usepackage{stmaryrd}

\usepackage[linktocpage,breaklinks]{hyperref}
\usepackage[usenames,dvipsnames]{xcolor}

\usepackage{tensor}
\usepackage{graphicx}
\usepackage{epsfig}
\usepackage{epstopdf}

\usepackage{natbib}
\usepackage{hyperref}

\hypersetup{colorlinks=true,
            citecolor=NavyBlue,
            linkcolor=NavyBlue,
            urlcolor=Mulberry}

\def\be{\begin{equation}}
\def\ee{\end{equation}}
\def\vp{\varphi}

\newcommand{\mmc}[1]{{\color{Red}{#1}}}
\begin{document}

\title{Cosmological attractors to general relativity \\
and spontaneous scalarization with disformal coupling}

\author{Hector O. Silva}
\affiliation{Department of Physics,
University of Illinois at Urbana-Champaign, Urbana, Illinois 61801, USA}
\affiliation{eXtreme Gravity Institute, Department of Physics,
Montana State University, Bozeman, Montana 59717, USA}

\author{Masato Minamitsuji}
\affiliation{Center for Astrophysics and Gravitation (CENTRA),
Instituto Superior T\'ecnico, University of Lisbon,
Lisbon 1049-001, Portugal.}

\begin{abstract}
The canonical scalar-tensor theory model which exhibits
spontaneous scalarization in the strong-gravity regime of
neutron stars has long been known to predict a cosmological
evolution for the scalar field which generically results in
severe violations of present-day Solar System constraints on
deviations from general relativity.
We study if this tension can be alleviated by generalizing
this model to include a disformal coupling between the
scalar field $\varphi$ and matter, where the Jordan frame
metric ${\tilde g}_{\mu\nu}$ is related to the Einstein
frame one $g_{\mu\nu}$  by ${\tilde g}_{\mu\nu}=A(\varphi)^2
(g_{\mu\nu}+\Lambda\, \partial_\mu \varphi \,
\partial_\nu\varphi)$.
We find that this broader theory admits a late-time
attractor mechanism towards general relativity. However, the existence of
this attractor requires a value of disformal scale of the
order $\Lambda\gtrsim H_0^{-2}$, where $H_0$ is the Hubble
parameter of today, which is much larger than the scale
relevant for spontaneous scalarization of neutron stars
$\Lambda \sim R_s^{2}$ with $R_s (\sim 10^{-22} H_0^{-1})$
being the typical radius of these stars.
The large values of $\Lambda$ necessary for the attractor mechanism (i)
suppress spontaneous scalarization altogether inside neutron stars and (ii)
induce ghost instabilities on scalar field fluctuations, thus preventing a
resolution of the tension.  We argue that the problem arises because our
disformal coupling involves a dimensionful parameter.
\end{abstract}

\maketitle

%%%%%%%%%%%%%%%%%%%%%%%%%
\section{Introduction}
\label{sec:introduction}
%%%%%%%%%%%%%%%%%%%%%%%%%

Einstein's theory of general relativity (GR) has passed all experimental tests
to date, ranging from the weak-field, low-velocity regime from of the Solar
System to the strong-field, low-velocity regime of binary
pulsars~\cite{Will:2014kxa}.
With the advent of gravitational-wave astronomy a new frontier for testing GR
has opened, providing us with the first glimpses of relativistic gravity in its
strong-field, high-velocity, nonlinear regime and the first direct probe into
the radiative properties of the theory~\cite{Berti:2015itd,Berti:2018cxi,Berti:2018vdi}.

To make the most out of this new arena for experimental gravity, it is
important not only to confront the predictions of GR against observations, but
also to embed it in a large theory space, obtained by relaxing one (or more) of
the fundamental pillars of GR and then letting experiments guide us towards the
region of this theory space which is most favorable by
observations~\cite{Damour:1996xx}.

In the vast landscape of extensions to GR, scalar-tensor theories stand out as
one of the simplest and most well motivated~\cite{Damour:1992we,Fujii:2003pa}.
In their simplest variant, they introduce a new scalar degree of freedom
($\vp$), violating the fundamental pillar of GR that gravity is mediated by a
single spin-2 field.
A simple scalar-tensor theory can be described (in the Einstein frame) by the
action
\begin{align}
S &= \frac{1}{2\kappa} \int d^4x \sqrt{-g}
\left(
R + 4X
\right)
+\int d^4x\sqrt{-\tilde g}\,
{\mathscr L}_{m}\left[{\tilde g}_{\mu\nu},\Psi\right],
\nonumber \\
\label{eq:e_ac}
\end{align}
where $g_{\mu\nu}$ and  ${\tilde g}_{\mu\nu}$ are
respectively the Einstein and Jordan frame metrics, $g \equiv {\rm det}
(g_{\mu\nu})$ and ${\tilde g} \equiv {\rm det} ({\tilde g}_{\mu\nu})$, and $R$
is the Ricci scalar curvature associated with $g_{\mu\nu}$,
$\kappa \equiv (8\pi G)/c^4$ where $G$ is the gravitational
constant in the Einstein frame and $c$ the speed of
light.
Finally, $X \equiv -(1/2) g^{\mu\nu} \vp_{\mu} \vp_{\nu}$, where
$\varphi_\mu\equiv \nabla_\mu \varphi$ is the covariant derivative of the
scalar field associated the metric $g_{\mu\nu}$ and
${\mathscr L}_{m}$ is the Lagrangian density of matter fields
$\Psi$ which couple minimally to $\tilde{g}_{\mu\nu}$.

In Ref.~\cite{Damour:1993hw}, it was shown that these theories
can not only pass Solar System constraints, but
also allow for large deviations relative to GR in the
strong-field regime found in neutron star (NS) interiors,
through a process known as {\it spontaneous
scalarization}.
In the simplest case where the two metrics are related by
a conformal transformation
\begin{equation}
{\tilde g}_{\mu\nu}= A(\vp)^2 g_{\mu\nu},
\end{equation}
the scalar field can become tachyonic unstable if $(\ln
A)_{,\varphi\varphi} < 0$, resulting in a NS which supports
a nontrivial scalar field configuration~\cite{Damour:1996ke,Harada:1998ge}.
For an exponential coupling $A(\vp)=\exp(\gamma_\alpha \vp^2/2)$,
spontaneous scalarization of static and spherically symmetric NSs
can happen below the threshold
$\gamma_\alpha\lesssim -4.35$ \cite{Harada:1997mr,Harada:1998ge},
depending weakly on the NS equation of state (EOS) and fluid
properties~\cite{Novak:1998rk,Silva:2014fca}.
On the experimental side, binary-pulsar observations (see
e.g.~\cite{Freire:2012mg,Archibald:2018oxs,Shao:2017gwu,Anderson:2019eay})
have placed the bound $\gamma_\alpha \gtrsim -4.5$.
These two results confine $\gamma_\alpha$ to a very
limited range, in which the effects of scalarization on
isolated NSs are bound to be small.

It was soon realized in~\cite{Damour:1992kf,Damour:1993id}
that the parameter space region in which the tachyonic
instability of the scalar field ($\gamma_\alpha<0$) can
happen for NSs would also affect the scalar field's
cosmological evolution, leading to large violations
of present-day Solar System constraints unless significant
fine-tuning is imposed at the time of matter-radiation
equality.
Conversely, when $\gamma_\alpha > 0$, the GR solution with
$\varphi = 0$ is an attractor of the theory, just after the
matter-radiation equality time, making the theory consistent
with present-day observations, but then preventing
spontaneous scalarization from happening.

While scalar-tensor theories which exhibit spontaneous
scalarization can still be used as toy models to explore
strong-field gravity phenomenology, ideally one would like
to find a model which reconciles its cosmology with
present-day physics.
Considerable effort has been placed on this issue recently.
For instance, Ref.~\cite{Anderson:2016aoi} considered
higher-order polynomial corrections to the quadratic
conformal coupling $\ln A =\gamma_\alpha \varphi^2/2+\delta
\varphi^4/4+\cdots$ (with $\delta>0$), where the
higher-order terms make it possible to satisfy the Solar
System constraints, but weakening considerably
scalarization.
Another possibility to solve this issue was presented in
Ref.~\cite{Anson:2019ebp} where, during inflation,
$\varphi$ gets a larger effective mass through a coupling to
the inflaton ($\psi$) of the form $g^2\psi^2\varphi^2$.
This coupling suppresses exponentially the amplitude of
$\varphi$ by the end of inflation and thus realizes the
otherwise \emph{ad hoc} fine-tuning previously mentioned.
Then, even if $\varphi$ grows after inflation, its
amplitude at present day could still be small enough to
satisfy Solar System constraints.

Here we explore whether this issue can be resolved by
introducing a disformal coupling between matter and
the scalar field.
More specifically, we consider a more general form for
$\tilde{g}_{\mu\nu}$ [appearing in Eq.~\eqref{eq:e_ac}], now
related with $g_{\mu\nu}$ by a disformal transformation,
\begin{equation}
\label{eq:disformal}
{\tilde g}_{\mu\nu}=A^2 (\vp)
\left[g_{\mu\nu}+ \Lambda B(\vp)^2\vp_{\mu}\vp_{\nu}\right],
\end{equation}
where $\Lambda$ is a constant with dimensions of $({\rm length})^2$.
Disformal transformations were originally introduced by
Bekenstein as the most general metric transformation
constructed from the metric $g_{\mu\nu}$ and the scalar
field $\varphi$ (and the first order derivative
$\varphi_{\mu}$) that respects causality and the weak
equivalence principle~\cite{Bekenstein:1992pj}. They have
been studied mainly in
cosmology~\cite{Koivisto:2008ak,Sakstein:2014aca,Sakstein:2015jca,Magueijo:2003gj,Kaloper:2003yf,vandeBruck:2015tna,Creminelli:2014wna,Minamitsuji:2014waa,Tsujikawa:2014uza,Watanabe:2015uqa,Motohashi:2015pra,Domenech:2015hka}
and have also been shown to allow for spontaneous
scalarization of NSs~\cite{Minamitsuji:2016hkk,Andreou:2019ikc}.
Modern scalar-tensor theories such as Horndeski
gravity~\cite{Horndeski:1974wa,Deffayet:2009wt,Kobayashi:2019hrl}
allow for conformal/disformal couplings to matter
fields~\cite{Zumalacarregui:2013pma,Bekenstein:1992pj,Bettoni:2013diz}
and they also preserve the mathematical
structure of the theory~\cite{Bettoni:2013diz}.

Is there any reason to expect that a disformal coupling
could remedy the issue outlined above?~Let us introduce the
functions which control the interaction strength between
scalar field and matter arising from the purely conformal
($A$) and purely disformal ($B$) terms of
Eq.~\eqref{eq:disformal},
\begin{equation}
\label{eq:alphabeta}
\alpha(\varphi) \equiv \frac{d\log A(\varphi)}{d\varphi},
\quad
\beta(\varphi) \equiv \frac{d\log B(\varphi)}{d\varphi}.
\end{equation}
The value of $\beta(\vp_0)$, where $\vp_0$ is the
cosmological value of the scalar field at the present time,
is poorly constrained~\cite{Ip:2015qsa}, because in the
nonrelativistic regime, where the pressure is negligible and
the scalar field is slowly varying relative to cosmological time
scales the disformal coupling becomes negligibly
small.\footnote{When the scalar field time dependence is negligible,
the disformal term contributes only past the second
post-Newtonian (PN) order and therefore does not affect the
parametrized post-Newtonian (PPN) parameters
$\gamma_{\rm PPN}$ and $\beta_{\rm PPN}$, which are identical
to those of `conformal' scalar-tensor gravity.}
However, since the scalar field $\vp$ varies on a
cosmological timescale, the disformal interaction is
expected to impact the cosmic expansion history, potentially
as important as the conformal contribution.
This opens the possibility that the disformal
interaction may quench the growth of the scalar
field in the regime $\gamma_\alpha < 0$ in which
scalarization happens~\cite{Minamitsuji:2016hkk} and at
the same time make the model consistent with Solar
System constraints.
Indeed, as we show later, the presence of the simplest disformal coupling $B=1$
is sufficient for the existence of a late-time attractor mechanism to GR, in
which $\varphi = 0$.
However,
the presence of the GR
attractor requires very large magnitudes of disformal
coupling $\Lambda<0$
so large that scalar field fluctuations suffer from ghost instability.

The existence of the late-time attractor mechanism can be
qualitatively understood as follows.
From Eq.~\eqref{eq:disformal}, assuming $\varphi\sim 1$, $B\sim 1$,
$\vp_\mu\sim \vp/R_s$ (with $R_s \sim 10$~km being the typical NS
radius) the disformal coupling could be as important as the conformal
coupling in NSs when $\Lambda\sim 100\, {\rm km}^2= 10^{12}\, {\rm cm}^2$
\cite{Minamitsuji:2016hkk}.
On the other hand, as our quantitative analysis shows,
the effective force which drives the cosmological evolution of the
scalar field in the presence of the disformal term
[see Eq.~\eqref{eff_force} for the precise definition]
is given by $-M_{\rm eff}^2\varphi$, where $M_{\rm eff}^2\sim
\gamma_\alpha H^2/(1+ k \Lambda H^2)$ is the effective mass of $\varphi$,
$k$ is a dimensionless constant of ${\cal O} (1)$, and $H$
is the Hubble expansion rate at the given moment of time.
Starting with an initial condition $\dot{\varphi}=0$ (where
$\dot{\varphi}$ is the derivative of $\varphi$ with respect to the
cosmological proper time) the amplitude of $\varphi$ remains constant
during the matter-dominated phase when $M_{\rm eff}/H\ll1$.
When $M_{\rm eff}/H \sim 1$, the effective force starts to
act of $\varphi$, driving the scalar field towards
$\varphi=0$ for $\gamma_\alpha<0$ as long as $\Lambda<0$.
The existence of the GR attractor for $\gamma_\alpha<0$
requires that $\varphi$ starts to feel the effective force
in the vicinity of present day, $M_{\rm eff,0}/H_0\sim
1$, where $H_0\sim 10^{-28}\, {\rm cm}^{-1}$ is the Hubble
parameter of today, and therefore $\Lambda\sim -H_0^{-2} \sim
-10^{56}\, {\rm cm}^2$.
Thus, the magnitude of $\Lambda$ which is necessary for the
existence of the GR attractor is larger than the $\Lambda$ for
scalarization of NSs by 44 order of magnitude, a
prohibitively large value for the theory to even allow for
the existence of scalarized relativistic
stars~\cite{Minamitsuji:2016hkk}.

In the rest of this work we present the details which
led to these conclusions.
In Sec.~\ref{sec:overview} we present the theory's field
equations and derive the equations which describe cosmology
in this theory.
In Sec.~\ref{sec:gr_sols} we study analytically the existence of
GR-attractor solutions when $\gamma_\alpha < 0$ and verify
their existence numerically in
Sec.~\ref{sec:num}, also relating our results with
spontaneous scalarization of NSs.
Finally, in Sec.~\ref{sec:conclusions} we present our
conclusions.
Hereafter we use geometrical units where $c=G=1$.

%%%%%%%%%%%%%%%%%%%%%%%%%
\section{Cosmological equations}
\label{sec:overview}
%%%%%%%%%%%%%%%%%%%%%%%%%

Let us start by describing the field equations of the theory
given by the action~\eqref{eq:e_ac} with the disformal
coupling~\eqref{eq:disformal}.
Variation of the action with respect to the Einstein frame
metric $g_{\mu\nu}$ results in the Einstein field equations
\begin{equation}
G^{\mu\nu}=\kappa \, \left(T_{(m)}^{\mu\nu}+ T_{(\varphi)}^{\mu\nu}\right),
\label{eq:eins}
\end{equation}
where the energy-momentum tensors of matter fields $\Psi$
and scalar field $\varphi$ are given by
\begin{align}
T_{(m)}^{\mu\nu}
\equiv
\frac{2}{\sqrt{-g}}
\frac{\delta \left(\sqrt{-\tilde g} {\cal L}_{m}\left[{\tilde g}(\varphi),\Psi\right]\right)}
{\delta g_{\mu\nu}},
\end{align}
and
\begin{align}
T_{(\varphi)}^{\mu\nu} &\equiv
\frac{4}{\kappa}
\frac{1}{\sqrt{-g}}
\frac{\delta \left(\sqrt{-g} \, X \right)}
{\delta g_{\mu\nu}}
\nonumber \\
&=
\frac{2}{\kappa}
\left(
\varphi^\mu\varphi^\nu
-\frac{1}{2} g^{\mu\nu}
\varphi^\alpha
\varphi_\alpha
\right),
\end{align}
respectively, where $\varphi^\mu \equiv g^{\mu\nu}\varphi_\nu$.

Variation of the action~\eqref{eq:e_ac} with
respect to $\varphi$ results in the scalar
field equation of motion
\begin{equation}
    \Box\varphi=(\kappa/2) {\cal Q},
\label{eq:scalar}
\end{equation}
where the function ${\cal Q}$ characterizes the strength of the coupling of matter
to the scalar field~\cite{Minamitsuji:2016hkk}
\begin{align}
\label{eq:coup}
{\cal Q} &\equiv
-\alpha (\varphi)
 T_{(m)}
+ \Lambda
\nabla_\rho
\left(
B(\varphi)^2
T_{(m)}^{\rho\sigma}
\varphi_\sigma
\right)
\nonumber\\
&\quad
-\Lambda B(\varphi)^2
\left[
\alpha(\varphi)
+\beta(\varphi)
\right]
 T_{(m)}^{\rho\sigma}
 \varphi_\rho
 \varphi_\sigma,
\end{align}
where $T_{(m)} \equiv g^{\rho\sigma}T_{(m)\rho\sigma}$ is the trace of
$T_{(m)\rho\sigma}$, and $\alpha(\varphi)$ and $\beta(\varphi)$ were defined
in Eq.~\eqref{eq:alphabeta}.
Observe that terms proportional to $\Lambda$ in~\eqref{eq:coup} are nonzero
even for the trivial choice $B=1$.
By taking the divergence of~\eqref{eq:eins}, employing the
contracted Bianchi identity $\nabla_\rho G^{\rho\sigma}=0$,
and using the scalar field equation of motion~\eqref{eq:scalar}, we obtain
\begin{equation}
\label{eq:cons}
\nabla_\rho T_{(m)}^{\rho\sigma}
=-\nabla_\rho T_{(\varphi)}^{\rho\sigma}
=-{\cal Q}\varphi^\sigma.
\end{equation}
Therefore, the coupling strength ${\cal Q}$ can be rewritten as
\begin{equation}
\label{step1}
{\cal Q}=
\Lambda B(\varphi)^2
\left(\nabla_\rho T_{(m)}^{\rho\sigma}\right) \varphi_\sigma
+{\cal Y},
\end{equation}
where we have introduced
\begin{align}
{\cal Y}
&\equiv
\Lambda B(\varphi)^2
\left\{
\left[
\beta (\varphi)
-\alpha (\varphi)
\right]
T_{(m)}^{\rho\sigma}
\varphi_\rho\varphi_\sigma
+
T_{(m)}^{\rho\sigma} \varphi_{\rho\sigma}
\right\}
\nonumber\\
&\quad -
\alpha(\varphi)
T_{(m)}.
\end{align}
Multiplying Eq.~\eqref{eq:cons} by $\varphi_\sigma$ and
solving it with respect to $(\nabla_\rho T_{(m)}^{\rho\sigma} )\varphi_\sigma$, we obtain
\begin{equation}
\label{eq:chi}
\chi
(
\nabla_\rho T^{\rho\sigma}_{(m)}
)\varphi_\sigma =2X{\cal Y},
\quad
\chi \equiv 1-2\Lambda B(\varphi)^2X.
\end{equation}
Then, substituting Eq.~\eqref{eq:chi} in~\eqref{step1}, using
$\cal{Q}={\cal{Y}}/{\chi}$, and finally eliminating
$\cal{Q}$ from~\eqref{eq:scalar}, we obtain the reduced
scalar field equation of motion
\begin{align}
\label{eq:scalar_eq}
\Box\varphi
&=
\frac{\kappa}{2\chi(X,\varphi)}
\left\llbracket
\Lambda B(\varphi)^2
\left\{
\left[
 \beta(\varphi) %\frac{B_\varphi}{B(\varphi)}
-\alpha(\varphi) %\frac{A_\varphi}{A(\varphi)}
\right]
T^{\rho\sigma}_{(m)}\varphi_\rho \varphi_\sigma
\right.
\right.
\nonumber\\
&\quad
\left.
\left.
+\,
T^{\rho\sigma}_{(m)}\varphi_{\rho\sigma}
\right\}
-\alpha (\varphi)
T_{(m)}
\right\rrbracket.
\end{align}

We consider the spatially flat
Friedmann-Lem\^eitre-Roberton-Walker (FLRW) spacetime in the
Einstein frame
\begin{equation}
ds^2
% g_{\mu\nu}dx^\mu dx^\nu
=-dt^2+a(t)^2\delta_{ij}dx^i dx^j,
\label{eq:efm}
\end{equation}
where $t$, $x^i$ are the coordinates of the time and the
three-dimensional space and assume that the scalar field
is only a function of time, i.e. $\varphi = \varphi(t)$~\cite{Damour:1992kf}.
The Jordan-frame metric is given by the FLRW line element
above by replacing the proper time $t \to {\tilde t}$ and
the scale factor $a \to {\tilde a}$.
These quantities are related as ${d\tilde t} \equiv A\sqrt{\chi} dt$
and ${\tilde a} \equiv Aa$.

We describe matter by a multicomponent perfect fluid,
with energy-momentum tensor in the Einstein and
Jordan frames denoted as
$T_{(m)}{}^{\mu}{}_{\nu}=\sum_a T_{(m)a}{}^{\mu}{}_{\nu}$
and
$\tilde{T}_{(m)}{}^{\mu}{}_{\nu}=\sum_a \tilde{T}_{(m)a}{}^{\mu}{}_{\nu}$,
respectively.
The fluid variables [pressure ($p$) and energy density
($\rho$)] in the two frames are related by
\begin{equation}
\label{rhoje}
{\tilde\rho}_a = \frac{\sqrt{\chi}}{A^4} \, \rho_a,
\quad
{\tilde p}_a = \frac{1}{A^4\sqrt{\chi}} \, p_a,
\end{equation}
where, from Eq. \eqref{eq:chi},
$\chi= 1-\Lambda B^2\dot{\varphi}^2$,
with an overdot denoting derivatives with respect
to $t$.
The EOS parameters of the
$(a)$th component of the fluid in the Jordan and Einstein
frames are defined by
${\tilde w}_a \equiv {\tilde p}_a/{\tilde\rho}_a$
and $w_a \equiv p_a/\rho_a$,
respectively, and are related by
\begin{equation}
w_a= \chi \, {\tilde w}_a.
\label{eos}
\end{equation}
Similarly, the EOS parameter for the whole
fluid is defined as
${\tilde w} = {\tilde p}/ {\tilde\rho}
= \sum_a {\tilde p}_a/\sum_a {\tilde\rho}_a$
and $w=p/\rho=\sum_a {p}_a/\sum_a \rho_a$,
which are also related by ${\tilde w} =w/\chi.$
The physically measured EOS parameter is that of the Jordan
frame and thus we should specify e.g. ${\tilde w}_a=0,
\, 1/3,\, -1$ to describe matter (i.e. dust), radiation,
and cosmological constant,
respectively.~Here, by cosmological constant, we also include the
equivalent vacuum energy.

Using Eq.~\eqref{eq:efm}, we find that the $(t,t)$-component
of the gravitational equations in the Einstein
frame~\eqref{eq:eins} reduces to
\begin{equation}
H^2
=\frac{\kappa}{3}\rho
+\frac{1}{3}\dot{\varphi}^2
=\frac{\kappa}{3}\sum_a\rho_a
+\frac{1}{3}\dot{\varphi}^2,
\label{ef1}
\end{equation}
where we have defined the Hubble parameter in the Einstein
frame $H \equiv \dot{a}/a$.
From Eq.~\eqref{eq:cons}, the energy conservation law of the
$(a)$th component yields
\begin{subequations}
\begin{align}
\label{ef3}
\dot{\rho}_a
&+3H
\left(
\rho_a+p_a
\right)
=\frac{{\cal Y}_a}{\chi}\dot{\varphi},
\\
\label{ef4}
\ddot{\varphi}
&+3H
\dot{\varphi}
% =-\frac{\kappa {\cal Y}}{2\chi},
=-\frac{\kappa}{2}\frac{{\cal Y}}{\chi},
\end{align}
\label{eq:fluid_scalar}
\end{subequations}
where ${\cal Y} = \sum_a {\cal Y}_{a}$ and
\begin{align}
{\cal Y}_a
&\equiv \Lambda B^2
\left[
(\beta-\alpha)\rho_a\dot{\varphi}^2
+\rho_a\ddot{\varphi}
-3H\dot{\varphi}\, p_a
\right]
\nonumber \\
&\quad +\alpha
(\rho_a-3p_a).
\end{align}

It is convenient to work with a rescaled time
coordinate $d\tau \equiv  H dt$~\cite{Damour:1992kf}, where $\tau=0$ corresponds
to the matter-radiation equality and $\tau=\tau_0$ denotes
the present day, which can be integrated as
$a(\tau)=a_0 \exp(\tau-\tau_0)$.
Hence $\tau$ describes the cosmic  $e$-folding time, where
$a_0$ is the size of the Universe today.
In terms of $\tau$, the Friedmann equation~\eqref{ef1}
becomes
\begin{equation}
\label{fjf}
H^2=\frac{\kappa \rho}{3-\varphi'^2}.
\end{equation}
We can then use Eqs.~\eqref{eos} and~\eqref{fjf} and
simplify Eqs.~\eqref{eq:fluid_scalar} (recast in terms of
$\tau$) to the forms
\begin{subequations}
\begin{align}
\label{eom1}
&\rho_a'+3\rho_a(1+\chi {\tilde w}_a)=\frac{{\cal Y}_a}{\chi}\varphi',
\\
\label{eom2}
&\left[
1+\frac{3}{2}
\lambda
 B^2\rho\frac{1-\varphi'^2}{3-\varphi'^2}
\right]
\frac{2\varphi''}{3-\varphi'^2}
\nonumber \\
&+
\biggl\{
1-\chi \tilde w
-\frac{
    \lambda B^2\rho}
        {2(3-\varphi'^2)}
\left[
3 (1+3\chi \tilde w)
+3(1-\chi \tilde w)
\varphi'^2
\right.
\nonumber \\
&\left.
+2(\alpha-\beta)\varphi'
\right]
\biggr\}\varphi'
=
-\alpha (1-3\chi \tilde w),
\end{align}
\label{eoms}
\end{subequations}
where $\lambda\equiv \kappa \Lambda$, and
\begin{align}
{\cal Y}_a
&=
\rho_a
\left\llbracket
\frac{
\lambda
 B^2 \rho}{3-\varphi'^2}
\left\{
 (\beta-\alpha)\varphi'^2
+\varphi''
\right. \right.
\nonumber \\
&\quad\left.\left. -\frac{1}{2}
\left[
3(1+3\chi {\tilde w}_a) +(1-\chi {\tilde w}_a)\varphi'^2
\right]
\varphi'
\right\}
\right.
\nonumber \\
&\quad \left.
+\, \alpha (1-3\chi \tilde w_a)
\right\rrbracket,
\end{align}
where primes indicate derivatives with respect to $\tau$
and now
\begin{equation}
\chi =1 - (\lambda \rho B^2 \varphi'^2)/(3-\varphi'^2).
\end{equation}
We note that in the radiation-dominated universe where
${\tilde w}_r=1/3$ and $\rho_r\gg \rho_a$ ($a\neq r$),
a nonzero constant constant scalar field $\varphi_{\ast}$
is a solution of the equations of motion
(see Sec.~\ref{sec:gr_sols}).
Equations~\eqref{eoms} are our main results from this section and
whose solutions are studied Secs.~\ref{sec:gr_sols}
and~\ref{sec:num}.
In the particular limit of purely conformal coupling
($\lambda=0$) these equations reduce to those
of Refs.~\cite{Damour:1992kf,Damour:1996ke}.

Before proceeding, we observe that
the Hubble parameters in both frames are related by
\begin{eqnarray}
{\tilde H}
=\frac{1+\alpha(\varphi) \varphi'}{A\sqrt{\chi}} H.
\end{eqnarray}
Using Eq. \eqref{rhoje},
the Friedmann equation Eq. \eqref{fjf} can be rewritten as
\begin{eqnarray}
{\tilde H}^2
=\frac{[1+\alpha(\varphi) \varphi']^2}{3-\varphi'^2}
 \frac{A(\varphi)^2}{\chi^{3/2}}
\kappa
{\tilde \rho}.
\end{eqnarray}
Using that Newton's constant in the Jordan frame at
present-day is
\begin{eqnarray}
G_{\rm eff}
=
% \frac{\kappa(1+\alpha(\varphi_0)^2) A (\varphi_0)^2}{8\pi},
[1+\alpha(\varphi_0)^2]\, A (\varphi_0)^2,
\end{eqnarray}
where $\varphi_0$ is the present day value of the scalar field,
the Friedmann equation can be written as
\begin{align}
{\tilde H}^2
=
\frac{3 [1+\alpha(\varphi) \varphi']^2}
       { (3-\varphi'^2)\chi^{3/2}}
\frac{1}{1+\alpha(\varphi_0)^2}
\left[
\frac{A(\varphi)}{A(\varphi_0)}
\right]^2
{ H}_{\rm GR}^2,
\end{align}
where ${H}_{\rm GR}$ is the expansion rate in the standard cosmology in GR.
Assuming that $\vp$ remains constant\footnote{In Sec.~\ref{sec:gr_sols} we show that an
arbitrary constant $\varphi$ is a solution in the
radiation-dominated era ${\tilde w}=1/3$ even in
the presence of the disformal coupling.} during big-bang
nucleosynthesis (BBN), say $\varphi=\vp_R$,
the ratio the
between Jordan-frame Hubble rates $\zeta(\tau)\equiv {\tilde H}/
{H_{\rm GR}}$ reduces to
\begin{equation}
\zeta(\tau_R) =
\frac{1}{\sqrt{1+\alpha(\varphi_0)^2}}
\frac{A(\varphi_R)}{A(\varphi_0)}.
\end{equation}
In order to be consistent with the observational BBN data,
$\zeta(\tau_R)$ has to satisfy $\left|1-\zeta(\tau_R)\right|
\leq 1/8$~\cite{Uzan:2010pm}, which combined with the Solar System constraint
$\alpha(\varphi_0)\ll 1$ gives
\begin{equation}
\label{bbn0}
\left|1-{A(\varphi_R)}/{A(\varphi_0)}\right|
\leq
{1}/{8}.
\end{equation}

%%%%%%%%%%%%%%%%%%%%%%%%%
\section{The GR attractor}
\label{sec:gr_sols}
%%%%%%%%%%%%%%%%%%%%%%%%%

So far we have worked with a general scalar-tensor theory,
keeping $A$ and $B$ as free functions. In this section, we
focus on a model which supports spontaneous scalarization
studied in~\cite{Minamitsuji:2016hkk}, consider
\begin{equation}
\label{exp}
A(\varphi)= e^{\gamma_\alpha\varphi^2 / 2},
\qquad
B(\varphi)= e^{\gamma_\beta\varphi^2 / 2},
\end{equation}
and examine under which conditions Eqs.~\eqref{eoms} admit a
cosmological GR attractor, which forces the scalar field to
evolve towards $\varphi = 0$.
As we see in this section, the choice of $\gamma_\beta$
does not affect the existence of the GR attractor and their
stability at all.

To gain some understanding on the existence of this
attractor, let us first consider the simplest case of a
single component of the fluid and a fixed scalar field
$\varphi=\varphi_{\ast}={\rm const.}$, in which
Eqs.~\eqref{eoms} reduce to
\begin{equation}
\rho= \rho_\ast e^{-3(1+\tilde w)\tau},
\qquad
\gamma_\alpha
\varphi_\ast
(1-3\tilde w)
= 0.
\end{equation}
For $\tilde w \neq 1/3$, the existence of the GR
solution requires $\varphi_\ast=0$, while for
$\tilde w= 1/3$ (i.e.  radiation) an arbitrary value of
$\varphi_\ast$ is a solution.

Since $\tilde w \neq 1/3$ in general, let us consider a small homogeneous
perturbation with respect to the GR solution $\rho=\rho_{\ast}
\exp[-3(1+\tilde w)\tau]+\delta\rho(\tau)$ and
$\varphi=\varphi_{\ast}+\delta\varphi(\tau)$.
Since $\delta\rho \propto \exp[-3(1+\tilde w)\tau]$, the
density perturbation behaves as the background solution and
can therefore be absorbed into it.
%%%
The perturbation for the scalar field satisfies
\begin{align}
\label{gr_pert}
&
\frac{1}{3}
\left(
2+
\lambda \rho_\ast e^{-3(1+\tilde w)\tau}
\right)
\delta \varphi''
\nonumber \\
&
+ \left[
1-\tilde w
% -\frac{\lambda}{2}\rho_\ast
-\rho_\ast\,(\lambda / 2)
%B_\ast^2
(1+3\tilde w)
  e^{-3(1+\tilde w)\tau}
\right]
\delta \varphi'
\nonumber \\
&+ \gamma_\alpha
\left(
1-3\tilde{w}
\right)
\delta\varphi
=0.
\end{align}
A late attractor to GR exists if the solution to
Eq.~\eqref{gr_pert} decays with time.

In the limit of a purely conformal coupling ($\lambda=0$), we
find that the solution to $\delta\varphi$ is given by
\begin{align}
\label{sol_conf}
\delta\varphi
\propto
{\rm exp}
\left[
-\frac{3\tau}{4}
\left(
1-\tilde w
\pm
\sqrt{
(1-\tilde w)^2
-\frac{8}{3}
\gamma_\alpha
(1-3\tilde w)
}
\right)
\right].
\nonumber \\
\end{align}
Assuming that $1-\tilde w>0$,
for $\gamma_\alpha(1-3\tilde w)>0$
both the solutions of Eq.~\eqref{sol_conf} decay with time $\tau>0$,
while for $\gamma_\alpha(1-3\tilde w)<0$
the `minus'-branch solution grows with time.
Thus, in the former case, the GR solution is an attractor.
For $\tilde w<1/3$, the condition for an GR attractor
reduces to $\gamma_\alpha>0$, consistent with the findings
of~\cite{Damour:1992kf,Damour:1993id}.

Now let us include the disformal coupling ($\lambda\neq 0$).
For $\tilde w>-1$, the disformal contribution decays
with time $\tau$, due to the exponentials appearing in
Eq.~\eqref{gr_pert}.
Hence, the disformal contribution is
negligible with respect to the conformal one, and $\delta\varphi$ can be
approximately given by Eq.~\eqref{sol_conf} at late times.
Consequently, the condition for the GR solution to be an attractor
is the same as in the purely conformal case.
On the other hand, for $\tilde w \leq -1$, the disformal
contribution is as important as the conformal one.
More specifically, for a cosmological constant ($\tilde
w=-1$) we have the equation
\begin{equation}
\left(
2+
%\kappa \Lambda
\lambda
 \rho_\ast
\right)
\left(
\delta \varphi''
+3\delta \varphi'
\right)
+12\gamma_\alpha
\delta\varphi
=0,
\end{equation}
which can be solved analytically:
\begin{equation}
\label{pert_sol_cc}
\delta \varphi \propto
{\rm exp}
\left[
-\frac{3\tau }{2}
\left(
1\pm \sqrt{1-\frac{4\bar{M}^2 }{9}}
\right)
\right],
\end{equation}
where $\bar{M}^2=12\gamma_\alpha/(2+\lambda \rho_\ast)$.
For $\bar{M}^2>0$, the solution decays with $\tau>0$, and
then the GR solution (i.e. a de Sitter Universe) is the
late-time attractor, if $\gamma_\alpha>0$ and $2+\lambda
\rho_\ast>0$, or if $\gamma_\alpha<0$ and $2+\lambda
\rho_\ast<0$.
Thus, in the latter case, the theory admits the existence of
the GR attractor even if $\gamma_\alpha<0$.
The stability of the GR attractor is discussed in
Appendix~\ref{app_b}.
We note that the negative sign of the
kinetic term signals the appearance of the ghost
instability.
However, we expect that during the matter-dominated phase
with the vanishing pressure $p_\ast=0$
both the gradient term and
the effective mass term
in the equation for the scalar field fluctuations \eqref{inh3}
are suppressed by the large factor
$|2+
\lambda \rho_\ast \exp[{-3(1+\tilde w)\tau}]|\gg 1$
(see Sec. \ref{sec:num}),
and hence
the growth of instability
would also be strongly suppressed and consequently
proceed slowly compared to the cosmological timescales.
During the dark energy (de Sitter) phase, both the gradient
and kinetic terms of the perturbations in the equation for
the scalar field fluctuations~\eqref{inh2} flip signs (see
Appendix~\ref{app_b}) and hence there would be no
exponential growth of the scalar field fluctuations.
However, the issue of the ghost instability may be significant at
present day and we will come back to it in
Sec.~\ref{sec:num}.

These conclusions can easily be extended for a
multicomponent fluid. For the constant scalar field,
$\varphi=\varphi_\ast$, the energy equation for the $(a)$th
component of the fluid and the scalar field equation of
motion are given by
\begin{equation}
\label{multi-gr}
\rho_a'+3 (1+w_a)\rho_a=0,
\qquad
\gamma_\alpha \varphi_\ast
\left(
1-3w
\right)
=0,
\end{equation}
where
\begin{equation}
1-3w
=\frac{\sum_a  (1-3w_a)\rho_a}
        {\sum_a\rho_a}.
\end{equation}
Thus, unless $\sum_a  (3w_a-1)\rho_a =0$ at all the
moments of time, GR solution is realized only for
$\varphi_\ast=0$.
However, in the radiation-dominated universe, the second
equation in Eq.~\eqref{multi-gr} can be approximately
satisfied for a constant scalar field $\varphi=\varphi_\ast\neq 0$.

If the
Universe is evolving towards the GR attractor, the
theory can (in principle) satisfy all the experimental
bounds on the parametrized post-Newtonian (PPN) parameters
of today~\cite{Will:2014kxa},
\begin{subequations}
\label{PPN_bounds}
\begin{align}
\gamma_{\rm PPN}&-1<2.3\times 10^{-5},
\\
\beta_{\rm PPN}&-1< 8\times 10^{-5},
\end{align}
\end{subequations}
where
\begin{subequations}
\begin{align}
\label{eq:a_a_ppn}
\gamma_{\rm PPN} &\equiv
\frac{1-\alpha(\varphi_0)^2}{1+\alpha(\varphi_0)^2},
\\
\beta_{\rm PPN}-1 &\equiv
\frac{\alpha_2(\varphi_0)}{2}
\frac{\alpha (\varphi_0)^2}{\left[1+\alpha (\varphi_0)^2\right]^2},
\label{eq:a_b_ppn}
\end{align}
\end{subequations}
and $\vp_0$, recall, is the present-day cosmological background value of the
scalar field. We also introduced
\begin{equation}
\alpha_2(\varphi_0)
\equiv
\frac{\partial^2\ln A(\varphi)}
     {\partial  \varphi^2}
    \Big|_{\varphi=\varphi_0}.
\label{eq:alpha_2}
\end{equation}
In deriving Eqs.~\eqref{eq:a_a_ppn} and \eqref{eq:a_b_ppn},
we have followed the standard procedure for calculating the PPN parameters
in our local Universe,
and ignored the cosmological time dependence of $\varphi$,
since the cosmological scalar field
varies with the cosmological timescale $10^{10}\, {\rm yr}$,
while the weak-field tests of gravity are done within
the light-crossing time in the Solar System $30\, {\rm au}/c
\sim 5\times 10^{-4}\,{\rm yr}$, where $30\,{\rm au}$ is the
approximated orbital radius of Neptune.
Thus, the corrections from the time dependence of
$\varphi$ are suppressed by some powers of the
ratio of these timescales.
Within these approximations,
corrections due to a nonzero disformal coupling appear
only via pressure effects, which are subdominant in the weak
gravity regime.

%%%%%%%%%%%%%%%%%%%%%%%%%
\section{Cosmological value of the scalar field and spontaneous scalarization}
\label{sec:num}
%%%%%%%%%%%%%%%%%%%%%%%%%

With intuition built on the existence of GR-attracted
solutions of Eqs.~\eqref{eoms}, we now numerically evolve
the scalar field in a realistic cosmology.

We assume that after inflation the cosmological expansion is
driven by the three components of the fluid in turns,
namely, radiation $\rho_r$, matter $\rho_m$, and the
cosmological constant $\rho_v$.

In the radiation-dominated phase just after inflation during
which $\rho_r\gg \rho_m,\rho_v$, the cosmological
expansion can be very well approximated by that
in the standard cosmology based on GR with a fixed amplitude
of the scalar field $\varphi_\ast$ which in general is
nonzero.
Strictly speaking, since even in the radiation-dominated
phase there is still very small contribution of nonrelativistic
particles, the force on the scalar field in the right-hand
side of Eq.~\eqref{eom2} does not vanish and
$\varphi$ would evolve in time very slowly.
Nevertheless, $\vp\approx \vp_\ast$ is a good approximation
during the radiation-dominated phase.

Next, $\rho_m$ eventually catches up with $\rho_r$ and
$\rho_m=\rho_r$ at the matter-radiation equality defined
to happen at $\tau = 0$.
As $\rho_m>\rho_r$, the scalar field $\varphi$ starts to
roll away from $\varphi=\varphi_\ast$ and the subsequent
dynamics requires the numerical integration of
Eqs.~\eqref{eoms}.
Since $\varphi={\cal O} (1)$ during most of the
evolution, as long as $\gamma_\beta={\cal O}(1)$ the
$\gamma_\beta$ dependence does not become significant for
cosmological dynamics.
%}
Thus, we set $\gamma_\beta=0$ in the rest of the paper,
although nonzero values
may be important in other contexts~\cite{Sakstein:2014isa,Sakstein:2014aca,Sakstein:2015jca,Minamitsuji:2016hkk}.

To do our numerical integration, we start from $\tau=0$ (the
matter-radiation equality) and we neglect $\rho_r$ in the
matter-dominated phase $\tau>0$, reducing our dynamical
variables to $\varphi$, $\rho_m$, and $\rho_v$.
From Eqs.~\eqref{eoms} we obtain the set of the evolution
equations,
\begin{align}
\label{eq_system}
&\rho_m'+3\rho_m=
% \frac{{\cal Y}_m}{\chi}\varphi',
({\cal Y}_m/\chi)\varphi',
\\
\label{eq_system2}
&\rho_v'+3\rho_v(1-\chi)=({\cal Y}_v/\chi)\varphi',
\\
\label{eq_system3}
&\left[
1+\frac{3\lambda\rho}{2}
\frac{1-\varphi'^2}{3-\varphi'^2}
\right]
\frac{2\varphi''}{3-\varphi'^2}
\nonumber \\
&+
\left\{
1-\chi \tilde w
-\frac{\lambda
\rho}
        {2(3-\varphi'^2)}
\left[
3 (1+3\chi \tilde w)
+3(1-\chi \tilde w)\varphi'^2
\right. \right.
\nonumber \\
&\left. \left.
 +2 \gamma_\alpha
\varphi\varphi'
\right]
\right\}\varphi'
=
-\gamma_\alpha\varphi
 (1-3\chi \tilde w),
\end{align}
where
\begin{align}
{\cal Y}_m
&=
\rho_m
\left\{
\frac{\lambda % e^{\gamma_\beta\varphi^2}
 \rho}{3-\varphi'^2}
 \left[
% (\gamma_\beta-\gamma_\alpha)\varphi\varphi'^2
-\gamma_\alpha \varphi\varphi'^2
+\varphi''
%\right. \right.
%\nonumber \\
%&\quad \left. \left.
-\frac{1}{2}
\left(
3 +\varphi'^2
\right)
\varphi'
\right]
\right.
% +\gamma_\alpha \varphi
\nonumber \\
&\quad \left. +\gamma_\alpha \varphi
\right\},
\\
{\cal Y}_v
&=
\rho_v
\left\llbracket
\frac{\lambda
%e^{\gamma_\beta\varphi^2}
 \rho}{3-\varphi'^2}
\left\{
% (\gamma_\beta-\gamma_\alpha)\varphi\varphi'^2
-\gamma_\alpha \varphi\varphi'^2
+\varphi''
\right. \right.
\nonumber \\
&\quad \left. \left.
-\frac{1}{2}
\left[
3(1-3\chi )
+(1+\chi)\varphi'^2
\right]
\varphi'
\right\}
+\gamma_\alpha (1+3\chi ) \varphi
\right\rrbracket,
\nonumber \\
\\
\chi
&=1-\frac{\lambda \rho  %e^{\gamma_\beta\varphi^2}
\varphi'^2}{3-\varphi'^2},
\end{align}
with
\begin{equation}
\rho=
\rho_m+\rho_v,
\qquad
{\tilde w}
=
\frac{\sum_a {\tilde w_a}  \rho_a}
        {\sum_a \rho_a}
=
-\frac{\rho_v}
        {\rho_m+\rho_v}.
\end{equation}
As initial conditions, we impose
\begin{align}
\label{initial}
\rho_m(0)&={\rho}_{m,e},
\quad
\rho_v (0)={\rho}_{v,e},
\\
\varphi(0)&=\varphi_0,
\qquad
\varphi'(0)=0,
\end{align}
where the subscript e denotes the quantities evaluated
at the matter-radiation equality.
It is convenient to identify an
effective force due to the disformal contribution
${\mathscr F}$ that acts
on $\vp$. This force is given by the right-hand side of
\eqref{eq_system3} divided by $1+(3\lambda/2)
(1-\varphi'^{2})/(3-\varphi'^2)$ i.e.
\begin{equation}
\label{eff_force}
{\mathscr F} \equiv -\frac{\gamma_\alpha
 (1-3\chi \tilde w)}
 {1+(3\lambda \rho/2)\,(1-\varphi'^2)(3-\varphi'^2)^{-1}}\,\varphi.
\end{equation}

In standard cosmology in which $\varphi=0$, matter
and radiation energy densities evolve according to $\rho_m (\tau)=\rho_{m,e} \exp(-3\tau)$
and $\rho_r(\tau)=\rho_{m,e} \exp(-4\tau)$,
where we have used the definition $\rho_{m,e}=\rho_{r,e}$.
We can then relate the proper time $\tau$ with ratio between
matter and radiation density as $\tau=\ln (\rho_m/\rho_r)$.
At present day ($\tau \equiv \tau_0$) the ratio
$\rho_m(\tau_0)/\rho_r (\tau_0)$ is approximately $3450$,
and hence $\tau_0= \ln (3450) \approx 8.15$.
Moreover, $\rho_{v,0} \approx 0.69 \rho_{\rm crit}$
and $\rho_{m,0} \approx 0.31 \rho_{{\rm crit}}$,
where $\rho_{\rm crit} (\approx 1.88\times 10^{-29} \, h^2 \, {\rm g / cm}^{3})$
is the critical energy density of today in standard
cosmology.
If the cosmological evolution follows that of standard
cosmology, we can rewrite the initial conditions
Eq.~\eqref{initial} as
\begin{align}
\label{initial2}
\rho_m(0)
&\approx
{\rho}_{m,0} e^{3\tau_0}
\approx
%\frac{0.31}{0.69} \rho_{0,\Lambda} e^{3\tau_0}
0.31 \,\rho_{\rm crit} e^{3\tau_0}
\approx
%1.84 \times 10^{10} \rho_{0,\Lambda},
1.27 \times 10^{10} \rho_{\rm crit},
\nonumber
\\
\rho_v(0)
&
\approx 0.69 {\rho}_{\rm crit},
\end{align}
which we also use for our integration in scalar-tensor theory.

In scalar-tensor cosmology, the ratio $\rho_m/\rho_r$
evolves differently from that in standard cosmology.
Since the
coupling between the scalar field and radiation is
negligible whenever $\varphi' \ll 1$, the evolution of
the radiation energy density follows closely that of the
standard cosmology $\rho_r(\tau)=\rho_{m,e}\exp(-4\tau)$.
On the other hand, the evolution of the matter energy
density $\rho_{m}(\tau)$ is in general nontrivial, even
in the presence of only the conformal coupling.
Thus, in the presence of the nontrivial conformal/disformal
couplings to the scalar field, matter and radiation
energy densities at the present day $\tau_0$ have to
satisfy
\begin{equation}
\frac{\rho_m(\tau_0)}
       {\rho_r (\tau_0)}
\approx \frac{\rho_m(\tau_0)}
       {\rho_{m,e} e^{-4\tau_0}}
=\frac{\rho_{m,0}}{\rho_{r,0}}
\approx 3450,
\end{equation}
which we use to define $\tau_0$ in scalar-tensor
cosmology.

If we assume that $\varphi_0\ll 1$ at the present day
[consistent with the bounds \eqref{PPN_bounds}]
and using the
identification of $\varphi_R=\varphi(0)$, the BBN constraint
\eqref{bbn0} can be rewritten as
$7/8 \leq \exp[(1/2)\gamma_\alpha \varphi(0)^2] \leq 9/8$.
For $\gamma_\alpha>0$ this yields
\begin{equation}
0 < \gamma_\alpha \varphi(0)^2 \lesssim  0.2355,
\label{bbn_initial1}
\end{equation}
while for $\gamma_\alpha<0$ we have
\begin{equation}
-0.2670 \lesssim  \gamma_\alpha \varphi(0)^2< 0,
\label{bbn_initial2}
\end{equation}
which can be used to fix a range of allowed scalar
field amplitudes $\varphi(0)$ consistent with BBN
constraints.

%%%%%%%%%%%%%%%%%%%%%%%%%%%%%%%
\begin{figure}[t]
\begin{center}
    \includegraphics[width=\columnwidth]{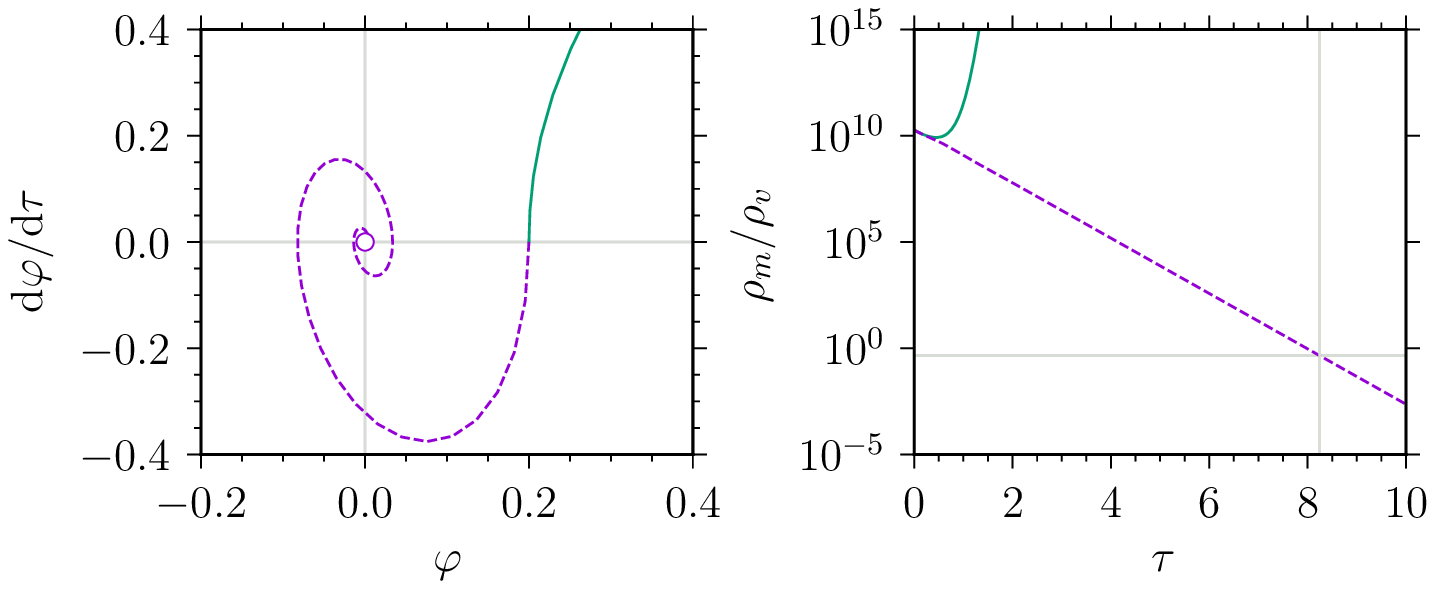}\\
    \includegraphics[width=\columnwidth]{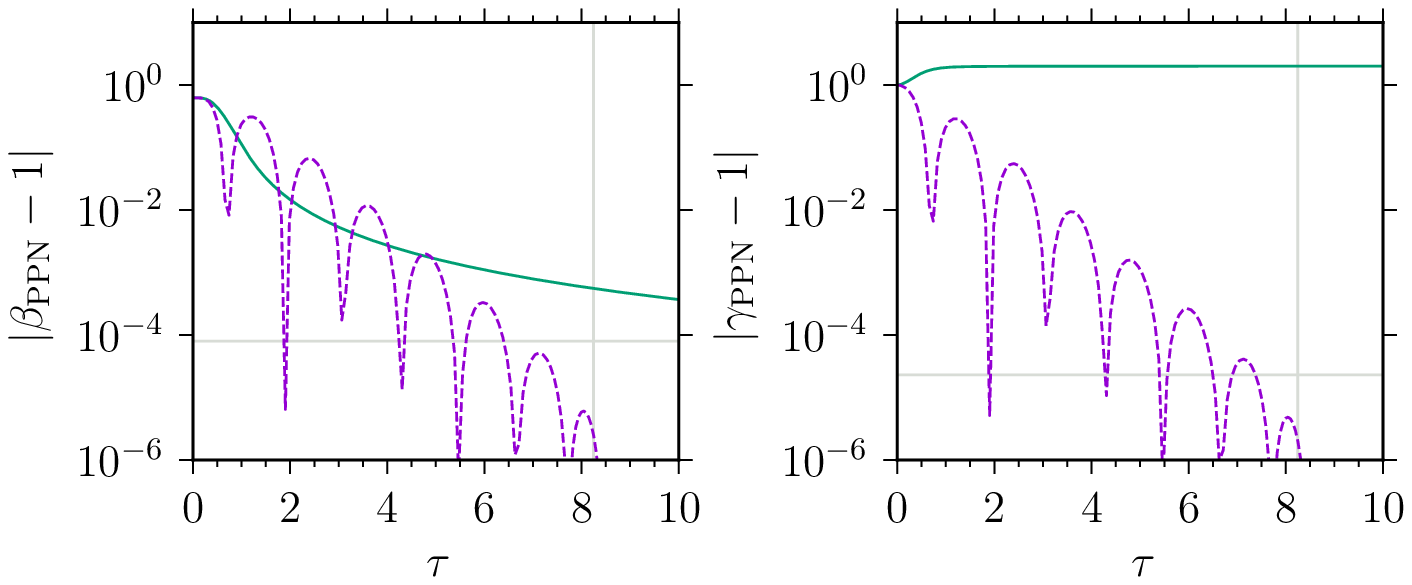}
\caption{
Scalar-tensor cosmology with only the conformal coupling.
In all panels, the solid curves correspond to $\gamma_\alpha = -5$,
while the dashed curves correspond to $\gamma_\alpha = 5$.
In both cases we used $\varphi(0) = 0.2$ as the initial
condition, which satisfies Eq.~\eqref{bbn_initial1}.
Top-left: The phase space portrait of the scalar
field's evolution (left panel) clearly shows the attractor
mechanism in action for $\gamma_\alpha > 0$.
For $\gamma_\alpha < 0$ the scalar field asymptotes to
infinity with constant `velocity' $\varphi' \approx 1.6$.
Top-right: the evolution of the ratio between matter
and cosmological constant densities. For $\gamma_\alpha >0$,
the present day $\rho_m / \rho_v = 0.455$
ratio happens around $\tau = 8.246$ which is
indicated by the circle in the top-left panel and by the
vertical lines in the other panels.
For $\gamma_\alpha < 0$, the ratio evolves to dramatically
violate observations.
Bottom row: the evolution of the PPN parameters
$\beta_{\rm PPN}$ (left) and $\gamma_{\rm PPN}$ (right). For
$\gamma_\alpha > 0$, the scalar field evolves as to satisfy the
PPN constraints~\eqref{PPN_bounds} (horizontal lines), while
for $\gamma_\alpha < 0$ both constraints are violated at
present day and in future.
}
\label{fig:conf}
\end{center}
\end{figure}
%%%%%%%%%%%%%%%%%%%%%%
%
Before studying the impact of the disformal coupling, we first
consider the case of the pure conformal coupling ($\lambda=0$).
In Fig.~\ref{fig:conf}, we show the results of integrating
the equations for $\gamma_\alpha=5$ (dashed curves) and
$\gamma_\alpha = -5$ (solid curves), using initial condition
$\varphi(0)=0.5$ which satisfies Eq.~\eqref{bbn_initial1} in
both examples.
In the top-left panel we show the phase space portrait of
$\varphi(\tau)$.
For $\gamma_\alpha < 0$, we see that the scalar field is
attracted towards %general relativity
GR  ($\varphi = 0$), while
for $\gamma_\alpha > 0$ the scalar field drifts away from
GR and asymptotes to infinity with constant
`velocity' $\approx 1.6$.
%
%In the
The top-right panel shows the evolution for $\rho_m/\rho_v$.
For $\gamma_\alpha < 0$, the present-day observed density
$\rho_m / \rho_v \approx 0.455$
% \mm{Please check this number, as $\rho_m/\rho_v$ should be around $0.45$ ?}
is reached at $\tau = 8.246$, which is indicated by the
circle in the top-right panel and the vertical lines in the
other panels.
For $\gamma_\alpha > 0$, $\rho_m / \rho_v$ evolves
inconsistently with observations.
The contrasting behavior of the scalar field, depending on
the sign of $\gamma_\alpha$, also reflects on the evolution
of the PPN parameters $\beta_{\rm PPN}$ and $\gamma_{\rm
PPN}$.
As shown in the bottom row, for $\gamma_\alpha < 0$ the value of
these parameters evolves towards being consistent with
present day PPN constraints~\eqref{PPN_bounds}, while for
$\gamma_\alpha > 0$ these constraints are strongly violated.
These results are consistent with those of~\cite{Damour:1992kf,Damour:1993id}.
% \mm{
% By ``Thus, for
% $\gamma_\alpha>0$ GR attractor at $\vp=0$ is reached
% during the matter-dominated phase'',
% I meant that the amplitude $\varphi$ stars to decrease toward zero
% during the matter-dominated phase.
% I agree that this sentence was misleading and should be removed,
% as $\varphi=0$ is never reached during this era.
% }

We now consider how the inclusion of the disformal coupling
changes this picture.
In the case of a pure disformal coupling
($\gamma_\alpha=0$), we observe that all the terms in the
scalar field equation of motion Eq.~\eqref{eq_system3} are
proportional to the derivatives $\varphi'$ or $\varphi''$.
Therefore, for the initial condition $\varphi'(0)=0$, $\vp$
remains constant, and consequently the cosmological
evolution will be the same as that in GR.

Now let us consider the more interesting case in which
both conformal and disformal terms contribute to the
scalar field dynamics. More specifically, we want to
examine if this case now admits an attractor to GR
when $\gamma_\alpha < 0$.
To do this, it is convenient to use the effective force
$\mathscr{F}$ defined in Eq. \eqref{eff_force}.
Since ${\tilde w}\leq 1$ and $|\varphi'|<1$,
we see that the effective force can be attractive
as long as $\lambda<0$ and  $|\lambda|\rho>1$ even when $\gamma_\alpha<0$.
When $|\lambda|\rho\gg 1$,
$\mathscr{F}$ becomes
of order ${\cal O}(\gamma_\alpha \varphi/(|\lambda|\rho))\ll \gamma_\alpha \varphi$,
and hence the effective force on the scalar field is
suppressed in comparison with the (pure) conformal case and
$\vp$ stays at the nearly constant amplitude $\vp(0)$.

When $|\lambda|\rho\sim 1$, the effective force starts to be
enhanced and $\vp$ is attracted towards $\vp=0$.
However, in the case that $|\lambda|\rho\sim 1$ is reached
during the matter dominated phase, since $\rho\propto\exp(-3\tau)$
decreases fast, the effective force term
Eq.~\eqref{eff_force} changes sign within the short period
and $\vp$ experiences a runaway growth after passing
through $\vp=0$.
On the other hand, in the case that $|\lambda|\rho\sim 1$ is
reached during the cosmological constant dominated phase,
the effective force term~\eqref{eff_force} does not change
sign, since $\lambda \rho$ is approximately constant and
$(1-\varphi'^2)(3-\varphi'^2)^{-1}$ varies only mildly,
without changing its sign.
Thus, the scalar field gradually approaches 0.
However, since at the present day $\rho_v \sim \rho_m$,
in general $\varphi(0) \sim 1$, which would easily be conflict with
the Solar System test.
Which of the two scenarios happens depends
on the magnitudes of $\lambda$, $\gamma_\alpha$ and the
value of $\varphi$ at the matter-radiation equality
time.
These imply that for a viable cosmology $|\lambda|\rho\sim
1$ has to be reached near the present day, when $\rho_v$
starts to catch up with $\rho_m$.
% \textcolor{red}{
Thus, it is convenient to normalize $\lambda$
in terms of the energy density of the cosmological constant
at the present day,
$\lambda=\lambda_x\equiv -10^x / (0.69 \rho_{\rm crit})$,
where $0<x<1$ is a constant parameter.
% }
%
%%%%%%%%%%%%%
\begin{figure}[t]
    \includegraphics[width=\columnwidth]{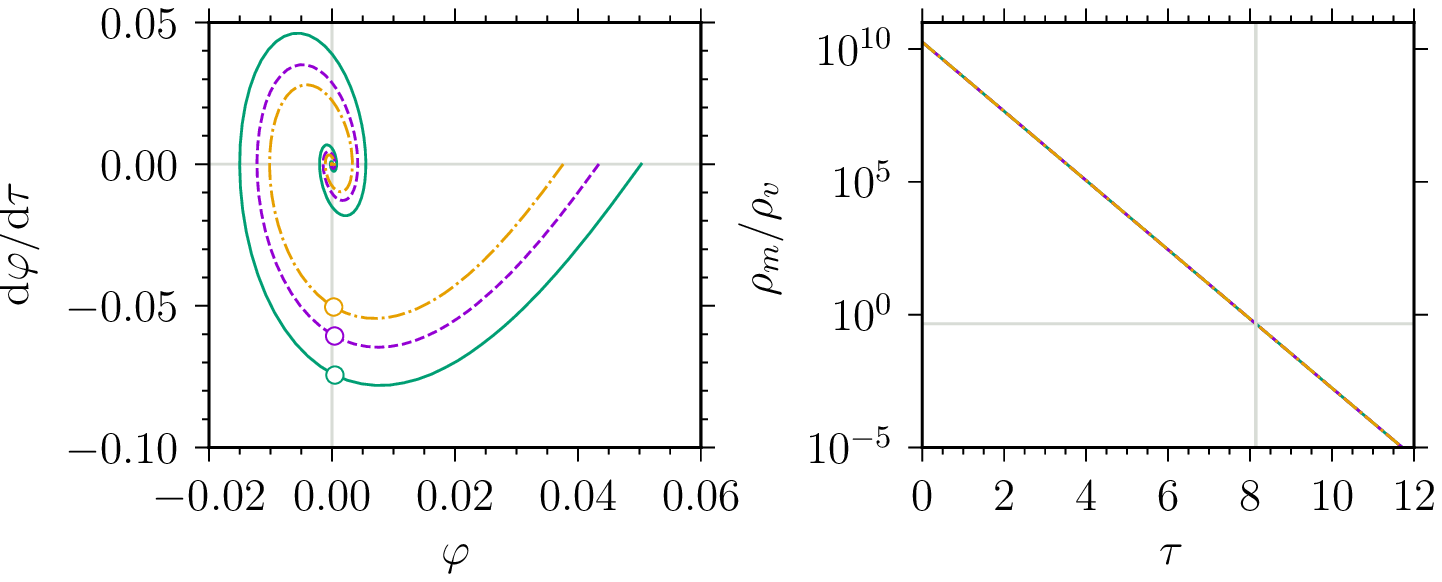}
    \includegraphics[width=\columnwidth]{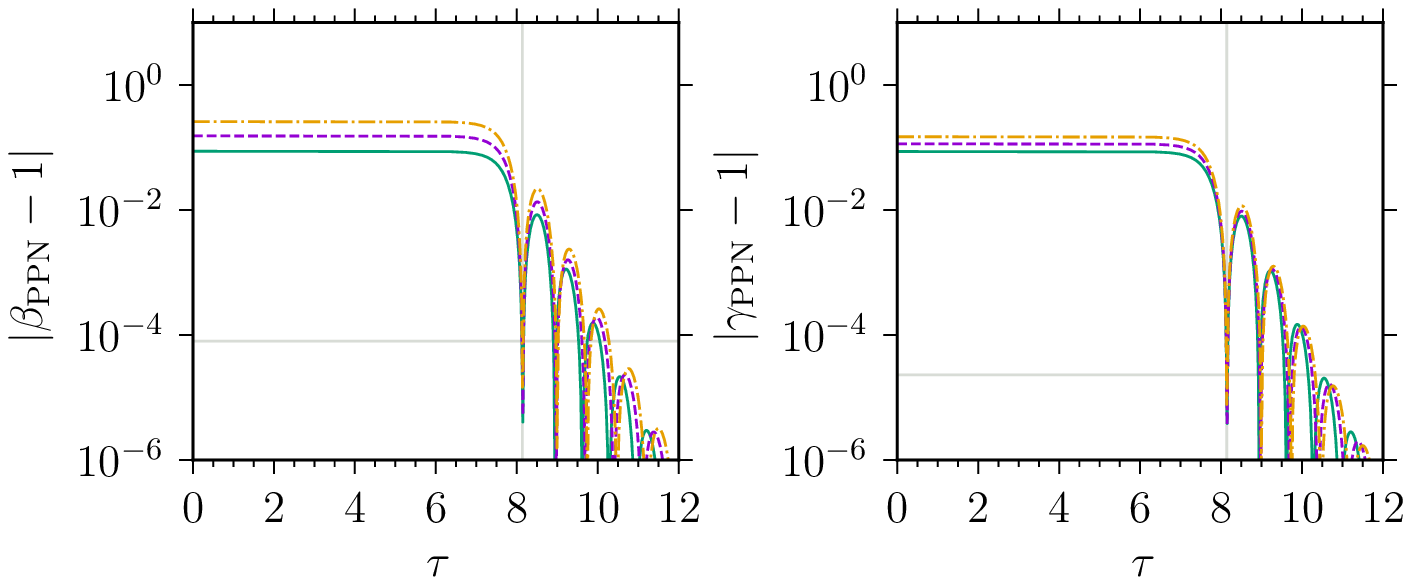}
\caption{
Scalar-tensor cosmology with disformal coupling.
In all panels, the solid curves correspond to $(\lambda_x, \gamma_\alpha) =
(\lambda_{0.6}, -4.22)$, the dashed curves to $(\lambda_{0.7}, -5.68)$ and the
dot-dashed curves to $(\lambda_{0.8}, -7.54)$.
We used the initial conditions $\vp(0)=0.0503, 0.0434, 0.0376$, respectively,
which satisfy the BBN constraint Eq.~\eqref{bbn_initial2}.
The panels are similar to those of Fig.~\eqref{fig:conf}, but here we focus
only on examples in which attractor mechanisms happen and therefore focus on
$\lambda < 0$.
Top-left: we show the phase space portrait of the scalar field. For
$\gamma_\alpha$ we see that the attraction towards GR persists, while it can
now also occur for $\gamma_\alpha < 0$.
Top-right: we show the ratio $\rho_m/\rho_v$. In all three cases they are
similar, visually indistinguishable. The present-day ratio $0.455$ is reached
at $\tau_0 \approx 8.14$ in all three examples.
The circles in the left panel and the horizontal line on the right panel are
the present-day $\tau_0\approx 8.14$, at which $\rho_{m}/\rho_{v} =0.455$.
}
\label{fig:dis}
\end{figure}
%%%%%%%%%%%%%%%%%%%%%%%%%%%%%%

In Fig.~\ref{fig:dis}, the left panels show the evolution of
$\varphi(\tau)$ and $\varphi'(\tau)$, while the right panels
show that of $\rho_m/\rho_v$.
The left panel shows the evolution of $\varphi(\tau)$ and
$\varphi'(\tau)$ in the phase space, and the right panel
shows that of $\rho_m/\rho_v$.  The solid, dashed, and
dot-dashed curves
correspond to the cases of
$(\lambda,\gamma_\alpha)=(\lambda_{0.6}, -4.22)$,
$(\lambda_{0.7}, -5.68)$,
and
$(\lambda_{0.8}, -7.54)$, respectively.
The corresponding initial conditions are $\vp(0)=0.0503$, $0.0434$, $0.0376$,
respectively, which satisfy the BBN constraint Eq.~\eqref{bbn_initial2}.
The circles in the left panel are the present-day
$\tau_0\approx 8.14$, at which
$\rho_{m}/\rho_{v} = 0.455$.
The PPN parameters at $\tau=\tau_0$ are given by
$(\gamma_{\rm PPN}-1,\beta_{\rm PPN}-1)\approx$
$(7.82\times 10^{-6}, -8.23\times 10^{-6})$,
$(1.26\times 10^{-5}, -1.78\times 10^{-5})$,
and
$(1.08\times 10^{-5}, -2.03\times 10^{-5})$,
respectively, which satisfy the PPN constraints~\eqref{PPN_bounds}.

The attractor mechanism to GR in this scalar-tensor theory
is reminiscent of the absence of spontaneous scalarization
of NSs in this theory when $\Lambda$ is negative and large
in magnitude~\cite{Minamitsuji:2016hkk}.
We have thus seen that the existence of a GR attractor and the compatibility
with the bounds on the PPN parameters requires that
$|\lambda|\rho_{\rm crit}\sim |\Lambda| (\kappa \rho_{\rm crit})\sim |\Lambda| H_0^2$,
$|\Lambda|\sim H_0^{-2}\sim 10^{56}\, {\rm cm}^2$.
What are the effects of such `disformal scale' on
gravitating systems?
Reference~\cite{Minamitsuji:2016hkk} (cf. Sec. VIII there) argued
that the kinetic part of the equation of motion for the
scalar field in the presence of a perfect fluid behaves as
\begin{equation}
- [1 - (|\lambda| / 2) \tilde{\rho}] \,\ddot{\vp},
\end{equation}
in a linearized approximation where $\chi \approx B \approx 1$.
Hence, assuming that $A\simeq 1$
and hence $\tilde\rho\simeq\rho$,
the kinetic term can flip sign (i.e.
cause a ghost instability) if
${\rho} \gtrsim 2 / |\lambda|$.
%($\equiv {\rho}_{\rm crit}$).
%
For the value $|\Lambda| \sim 10^{56}\,{\rm cm}^{2}$,
this implies a threshold
density
${\rho}_{t}
\sim 10^{-29}\,{\rm g/cm}^{3}$
necessary to induce the instability.
Moreover, the assumption $\chi \approx 1$ imposes $X \ll 1$ on the
scalar field's kinetic energy.
Therefore, this tremendously small density (of the same order of
magnitude as the cosmic mean density) indicates that
even though the Universe may be consistent with GR,
small fluctuations of the permeating scalar field
would necessarily be unstable.
We note that
in higher density regions
ghost instabilities
would proceed more slowly
due to
the presence of a larger coefficient
$|1-(|\lambda|/2){\tilde\rho}|$,
and hence
the instability may be more significant on larger length scales.

%%%%%%%%%%%%%%%%%%%%%%%%%
\section{Discussions}
\label{sec:conclusions}
%%%%%%%%%%%%%%%%%%%%%%%%%

We investigated whether in the presence of disformal
coupling scalar-tensor theories with conformal coupling
$\gamma_\alpha<0$ allows the GR attractor in the late-time
Universe, and if it is the case, whether the same coupling
is compatible with spontaneous scalarization of NSs.

We showed that the effect of disformal coupling could make
it possible to realize the GR attractor.
The effective force on the cosmological scalar field is
given by Eq.~\eqref{eff_force}. Even if $\gamma_\alpha<0$,
the effective force becomes attractive, if $\lambda<0$ and
$|\lambda|\rho>1$.
As long as $|\lambda|\rho\gg 1$ the effective force is
suppressed compared to the case of the pure conformal
coupling and $\varphi$ remains a nonzero constant, and when
the energy density of the Universe becomes lower as such
$|\lambda|\rho\sim 1$ the effective force starts to act and
$\varphi$ is attracted towards 0.
In the case $|\lambda|\rho\sim 1$ during the matter-dominated phase, since
$\rho$ exponentially decreases with respect to $\tau$ the force becomes
repulsive again when $\varphi$ approaches zero, and $\varphi$ grows again.
On the other hand, in the case $|\lambda|\rho \sim 1$ during the
cosmological constant-dominated phase, $|\lambda|\rho$ approaches a
constant value, while $\varphi$ approaches zero after oscillating
through $\varphi = 0$..
However, since the GR attractor is reached in the future,
the cosmological values of $\varphi$ could satisfy the PPN
constraints \eqref{PPN_bounds} unless the values of
couplings and/or initial conditions are fine-tuned.
Examples satisfying the bounds on the PPN parameters are
shown in Fig.~\ref{fig:dis}.

The disformal coupling which is necessary for the existence
of the GR attractor is given by
$\lambda \rho_0\sim \Lambda H_0^2\gtrsim 1$,
and hence $\Lambda\gtrsim H_0^{-2}$.
On the other hand, for spontaneous scalarization of NSs, the
typical value of the disformal coupling is given by
$\Lambda\gtrsim R_s^{2}$, where $R_s$ is a typical radius of
NSs.
Since $H_0^{-1}\sim 10^{22}\,R_s$, the value of $\Lambda$
necessary for the existence of the GR attractor is much
larger than that of spontaneous scalarization of NSs.
As argued in Ref.~\cite{Minamitsuji:2016hkk}, such a huge
value of disformal coupling prevents scalarization of
NSs and even worse, induce ghost instabilities
of matter present in all scales of the Universe.
Therefore, introducing a disformal coupling
does not help reconcile the spontaneous
scalarization model of~\cite{Damour:1993hw} with
cosmological evolution of the scalar field.
We expect that the problem is ubiquitous to any model with
spontaneous scalarization induced by dimensionful coupling
constant.
Such a large disformal coupling parameter $\Lambda$
might also affect local gravitational physics
and modify the expression of the leading-order
PPN parameters~\eqref{eq:a_a_ppn} and \eqref{eq:a_b_ppn}.
Even if there would be a change of the PPN parameters,
the Solar System constraints would be satisfied
only for the particular initial conditions
and our main results would not be affected.

Ultimately, the problem arises because $\Lambda$ is a
dimensionful coupling and hence the effective dimensionless
coupling crucially depends on the environment.
A conceptually similar problem was argued in the context of
embedding the model of black hole BH scalarization
of~\cite{Doneva:2017bvd,Silva:2017uqg} into the inflationary
cosmology~\cite{Anson:2019uto}, which involves a coupling to
the Gauss-Bonnet term
$\lambda^2 \varphi^2 (R^2-4R^{\alpha\beta}R_{\alpha\beta}
+R^{\alpha\beta\mu\nu}R_{\alpha\beta\mu\nu})$,
where the coupling $\lambda$ has dimension of $({\rm length})$.
In order to scalarize a BH with mass of $M={\cal
O}(M_\odot)$, where $M_{\odot}$ is the Solar mass,
the coupling has to be
$\lambda\sim GM \sim  M_\odot/M_{\rm Pl}^2\sim 10^{19}\,{\rm GeV}^{-1}$.
Assuming that the scalar field $\varphi$ is present at the
beginning of inflation, it is quantized in a Bunch-Davies
vacuum as the inflaton.
It was suggested that for $\lambda>0$
the same coupling induces a catastrophic production of the
$\varphi$-particles within the timescale
$(\lambda H_{\rm inf}^2)^{-1}\sim 10^{-32} (H_{\rm inf})^{-1}$,
assuming that the Hubble rate during inflation is given by
$H_{\rm inf}=10^{13}\,{\rm GeV}$.
Thus, quantum fluctuations of $\varphi$ would rapidly grow
and completely destroy the inflationary universe within the
timescale much smaller than the Hubble time.
This comes from the huge hierarchy between the two different
curvature lengths $GM\sim 10^{19}\,{\rm GeV}^{-1}$ and
$H_{\rm inf}^{-1} \sim 10^{-13}\,{\rm GeV}^{-1}$.
In our case, a similar problem arises from the huge hierarchy
between $R_s\sim 10^{6}\,{\rm cm}$ and $H_{0}^{-1} \sim 10^{28}\, {\rm cm}$.

At last, let us briefly comment on some possible extensions
of our work and also place our results in perspective with other
recent work of spontaneous scalarization.
First, in the context of scalar-tensor theories,
the disformal transformation~\eqref{eq:disformal} could be generalized by
the inclusion of a $X$-dependence, i.e.,
$A=A(X,\varphi)$ and $B=B(X,\varphi)$~\cite{Bekenstein:1992pj}.
This generalization maps the Lagrangian
in~\eqref{eq:e_ac} (after going to the Jordan frame)
% rewritten in terms of the Jordan frame metric ${\tilde g}_{\mu\nu}$
to a subclass of degenerate higher-order
scalar-tensor theories~\cite{Langlois:2015cwa,Achour:2016rkg,BenAchour:2016fzp}.
How this generalized disformal coupling influences
spontaneous scalarization of stars has not been investigated
yet and it would be interesting to perform an analysis
similar to that presented here for the cosmological
evolution of the scalar field.
Second, Ref.~\cite{Andreou:2019ikc} recently
isolated all the terms within Horndeski gravity
which can potentially induce a tachyonic instability at the
linear level (see also~\cite{Minamitsuji:2019iwp}).
They are the original Damour-Esposito-Far\`ese model~\cite{Damour:1993hw}, the
scalar-Gauss-Bonnet
theory~\cite{Antoniou:2017acq,Silva:2017uqg,Doneva:2017bvd}, and
the model with disformal coupling to matter (related to~\cite{Minamitsuji:2016hkk}).
A potential term of the scalar field, which cannot trigger
scalarization on its own, can however influence the
onset of the instability caused by the other three
terms.
Individually, each of the three `instability trigger'
terms have been shown to generically lead
to violations of Solar System constraints, while the
mass term can alleviate the tension depending on the
scalar field's mass~\cite{Alby:2017dzl}.
It would be interesting to study the cosmology of the full theory,
combining these three terms and study if the combination of more than one dimensionful
coupling parameters (e.g. arising from the scalar field's coupling to the
Gauss-Bonnet term and disformally to matter) could resolve the tension.
Moreover, other terms belonging to the Horndeski action
(or beyond Horndeski for more general models which satisfy the
recent bounds on the speed of gravitational waves
\cite{Monitor:2017mdv,Creminelli:2017sry,Ezquiaga:2017ekz,Baker:2017hug,Sakstein:2017xjx,Crisostomi:2017lbg,Langlois:2017dyl})
besides these four could be relevant for cosmological
evolution at the \textit{nonlinear} level, and could be able
to make spontaneous scalarization compatible with cosmology.

\acknowledgments{
It is a pleasure to thank Eugeny Babichev and Jeremy Sakstein for
comments and suggestions on this work.
We also thank Nicol\'as Yunes for discussions during the
development of this work.
H.O.S~was supported by the NASA Grants No.~NNX16AB98G and
No.~80NSSC17M0041.
M.M.~was supported by the research grant under the Decree-Law 57/2016 of
August 29 (Portugal) through the Funda\c{c}\~{a}o para a Ci\^encia e a
Tecnologia..
}

%%%%%%%%%%%%%%%%%%%%%%%%%
\appendix

%%%%%%%%%%%%%%%%%%%%%%%%%
\section{Stability of the GR attractor}
\label{app_b}
%%%%%%%%%%%%%%%%%%%%%%%%%

In this appendix, we briefly comment on the stability of the GR solution
against inhomogeneous perturbation of the scalar field $\varphi = \delta
\varphi(t,x^i)$, which follows from the equation
\begin{align}
\label{inh}
&\left(
2+\lambda {\tilde\rho}
\right)
\delta \ddot{\varphi}
+
\left(
2-\lambda {\tilde \rho}{\tilde w}
\right)
\left(
3H\delta\dot{\varphi}
-a^{-2} \Delta \delta\varphi
\right)
\nonumber\\
&+
\kappa \gamma_\alpha{\tilde\rho}
\left(1-3\tilde{w}\right)
\delta\varphi
=0\mmc{,}
\end{align}
where
$\Delta \equiv \delta^{ij} {\partial}_i {\partial}_j$
is the Laplacian operator.

For $2+\lambda  {\tilde\rho}<0$ and
$2-\lambda {\tilde \rho}{\tilde w}>0$,
$\delta\varphi$ suffers the ghost instability,
while for
$2+\lambda {\tilde\rho}>0$
and
$2-\lambda {\tilde \rho}{\tilde w}<0$,
$\delta\varphi$ suffers the spatial gradient instability.
On the other hand,
for $2+\lambda  {\tilde\rho}>0$
and
$2-\lambda {\tilde \rho}{\tilde w}>0$,
or for
$2+\lambda {\tilde\rho}<0$
and
$2-\lambda  {\tilde \rho}{\tilde w}<0$,
$\delta\varphi$ does not suffer any instability
arising from the modified kinetic term.
In the case of the cosmological constant $\tilde w=-1$,
Eq.~\eqref{inh} reduces to
\begin{equation}
\label{inh2}
\left(
2+\lambda  {\tilde\rho}
\right)
\left(
\delta \ddot{\varphi}
+3H\delta\dot{\varphi}
-a^{-2} \Delta \delta\varphi
\right)
+4\kappa \gamma_\alpha{\tilde\rho}
\delta\varphi
=0.
\end{equation}
Since the coefficients for the second derivative terms are
common, no ghost and gradient instability happen.
On the other hand, for matter $\tilde w=0$,
\begin{equation}
\label{inh3}
\left(
2+\lambda {\tilde\rho}
\right)
\delta \ddot{\varphi}
+2
\left(
3H\delta\dot{\varphi}
% -\frac{\Delta}{a^2} \delta\varphi
- a^{-2} \Delta\delta\varphi
\right)
+\kappa \gamma_\alpha{\tilde\rho}
\delta\varphi
=0.
\end{equation}
When $2+\lambda {\tilde\rho}<0$, the ghost instability
happens.
We note that for the intermediate non-GR solutions nonzero
$\dot{\varphi}$ and $\ddot{\varphi}$ would nontrivial
contribute to the kinetic terms of cosmological
perturbations, and the appearance of the ghost mode is
unclear.

\bibliographystyle{apsrev4-1}
\bibliography{disformal_refs}

\end{document}